\documentstyle[prl,aps]{revtex}
\draft

\begin{document}

\twocolumn

\title{Anderson transition in three-dimensional disordered systems with
symplectic symmetry}
\author{Tohru Kawarabayashi}
\address{Institute for Solid State Physics, University of Tokyo,
Roppongi, Minato-ku, Tokyo 106, Japan}
\author{Tomi Ohtsuki}
\address{Department of Physics, Sophia University,
Kioi-cho 7-1, Chiyoda-ku, Tokyo 102, Japan}
\author{Keith Slevin}
\address{The Institute of Physical and Chemical Research, Hirosawa 2-1,
Wako-shi, Saitama 351-01, Japan}
\author{Yoshiyuki Ono}
\address{Department of Physics, Toho University, Miyama 2-2-1, Funabashi,
Chiba 274, Japan}

\date{\today}
\maketitle

\begin{abstract}
The Anderson transition in a 3D system with
symplectic symmetry is investigated numerically. From a
one-parameter scaling analysis the critical
exponent $\nu$ of the localization length is extracted and
estimated to be
$\nu = 1.3 \pm 0.2$.
The level statistics at the critical point are
also analyzed and shown to be scale
independent.
The form of the energy level spacing distribution $P(s)$
at the critical point is found
to
be different from
that for the orthogonal ensemble suggesting that the
breaking of spin rotation symmetry is relevant at the
critical point.

\end{abstract}

\pacs{71.30.+h, 71.55.J, 72.15.Rn, 64.60.-i}

\narrowtext

Since the original work by Anderson \cite{Anderson} there
has been considerable interest in the metal-insulator transition in
disordered electron systems
\cite{LR,KM}.
Critical phenomena at the Anderson transition are
conventionally classified according to the
universality class into which the system falls: orthogonal, unitary
or symplectic \cite{Dyson,HLN}.
The Anderson transition has been studied intensively
numerically, analytically and experimentally.
Nevertheless, in spite of this effort, we believe it
is fair to say
that some aspects of the Anderson transition
remain puzzling. For example, the critical exponent $\nu$
of the localization length has been estimated for three-dimensional (3D)
systems with orthogonal \cite{KM} and unitary symmetry
\cite{OKO,HKO,CD} by means of
finite-size scaling. The estimated numerical values of the
critical exponent for these two universality classes have turned out
to be rather close to
each other.
At present it is unclear whether or not this is
an accidental coincidence.
An obvious question immediately presents itself; does
this coincidence also occur for the symplectic
symmetry class?
And if so does this coincidence also hold for
other characteristics of the critical point?

Systems belonging to the symplectic universality class exhibit
the Anderson transition even
in two-dimensions (2D)
\cite{EZ,Ando,FABHRSWW,KO}, while
systems belonging to the unitary and orthogonal classes
do not, in general,
exhibit any Anderson transition in 2D.
Thus we also have the opportunity, when studying symplectic
systems, to see how
the critical behavior at the Anderson transition
depends on the dimensionality of the system.

The statistical properties of energy levels \cite{Dyson,DM,Mehta}
in the vicinity of the Anderson transition,
and at the critical point in particular,
have attracted
considerable attention recently
\cite{AZKS,SSSLS,HS_3DO,OO_0,Evangelou_3DO,ZK,SZ,OO,Evangelou_2DS,HS}.
On the metallic side of the transition
it has been demonstrated that
the energy level statistics can
be described by random matrix
theory \cite{Dyson,DM,Mehta}; for example, the
spacing $s$ between successive energy levels is well described by a
distribution function $P(s)$ which is quite close to
the Wigner surmise.
In contrast, on the insulating side of the transition,
energy levels are uncorrelated and
the spacing distribution is Poissonian.
At the critical point, where the metal-insulator transition takes place,
it has been claimed \cite{AZKS,SSSLS}
that the energy level statistics are also universal
but different from those
predicted by random matrix theory. These so called
critical statistics are believed to be universal in the sense that
they depend neither on
the system-size nor on the details of the model Hamiltonian.
It is expected that these critical statistics should be determined
only by the symmetry of the system and
that they should be reflected in
the critical behavior at the Anderson transition \cite{KLAA,AKL}.
The existence of   universal critical statistics
has been demonstrated numerically for a 3D orthogonal system and
for several 2D systems
\cite{HS_3DO,SZ,OO,Evangelou_2DS,VHSP}.

For 2D symplectic systems, it has been
observed that the level
spacing distribution at the critical point $(P_{\rm c}(s))$ exhibits
a power-law behavior as $P_{\rm c}(s) \sim s^4$ for $s \ll 1$
\cite{SZ,OO,Evangelou_2DS}.
For 2D unitary systems it has been found
that the spacing distribution behaves
as $P_{\rm c}(s) \sim s^2$ for $s \ll 1$ \cite{OO}.
These behaviors of the level correlations
for small $s$ are consistent with
random matrix theory.
However, it has also been reported
\cite{HS} that in 3D $P_{\rm c}(s)$ is unaffected by the presence of
an Aharonov-Bohm flux (AB flux), even though the
AB flux should break time reversal symmetry.
If, as is claimed in \cite{HS}, $P_{\rm c}(s)$ is insensitive
to the breaking of time reversal symmetry it
immediately prompts the question; is $P_{\rm c}(s)$ also
insensitive to the breaking of spin rotation symmetry?

In this letter, we consider a 3D system with
symplectic symmetry. In particular, we estimate the
critical exponent $\nu$ for
the localization length and examine the spacing distribution
$P_{\rm c}(s)$ of the energy levels in the critical regime.

The model we adopt here is described by the Hamiltonian \cite{Keith}
\begin{equation}
 H = \sum_{<i,j>,\sigma,\sigma'}
 V_{i,\sigma;j,\sigma'}C^{\dagger}_{i,\sigma}
 C_{j,\sigma'} + \sum_{i,\sigma} \varepsilon_i C^{\dagger}_{i,\sigma}
 C_{i,\sigma}
\end{equation}
with
\begin{equation}
 V_{i,\sigma;i- k,\sigma'} =  V [\exp( -{\rm i}\theta
 \sigma_{k}) ]_{\sigma,\sigma'}, \qquad k = \hat{x}, \hat{y}
 , \hat{z}, \label{coupling}
\end{equation}
where $C^{\dagger}_{i,\sigma}(C_{i,\sigma})$ denotes a creation
(annihilation) operator of an electron at
the site $i$ with spin $\sigma$, and
$\{ \sigma_x, \sigma_y, \sigma_z \}$
denote the Pauli matrices.
The lattice sites $i$ are supposed to lie on a simple cubic
lattice.
The strength of the hopping amplitude is
denoted by $V$.
The site-diagonal potentials $\{ \varepsilon_i \}$ are assumed to be
distributed independently, and their distribution
is taken to be uniform in the range $[-W/2,W/2]$.
The parameter $\theta$ characterizes the strength of the
spin-orbit coupling. For $\theta = 0$ orthogonal symmetry is
recovered. Here, two cases of strong spin-orbit coupling,
$\theta = \pi/6$
and $\theta = \pi /4$, are studied.
The length is scaled by the lattice spacing and
$\hat{x}(\hat{y},\hat{z})$ denotes the unit vector in the
$x$-($y$-,$z$-)direction.
We note that this
model is a 3D generalization of the Ando model \cite{Ando}
proposed for a 2D symplectic system.
It is easy to verify that the Hamiltonian
is invariant under time reversal.
In the following, we confine ourselves to the case of the band center
$(E=0)$, for simplicity.

First, we carry out a finite-size scaling analysis \cite{MK}
of the Anderson transition
and estimate the critical exponent $\nu$.
We consider a quasi-one dimensional system $(M\times M \times
L)$ with $L \gg M$
whose cross section is composed of an $M \times M$ two-dimensional
lattice with a periodic boundary condition.
We calculate the localization length $\xi_M$ along such a
quasi-one dimensional system with
the transfer matrix method, in which we adopt the quaternion-real
representation
of the Hamiltonian so as to carry out
numerical multiplications efficiently.
The system-sizes treated here are $M=6,8,10$
and $12$
with $L$ up to $5\times 10^4$. An average over
four independent realizations of random potentials
has been performed in order to reduce
the error of raw data, resulting in relative errors smaller
than $1\%$.
In Fig. 1, the renormalized localization length $\Lambda_M$
defined by $\Lambda_M \equiv \xi_M/M $
is plotted as a function of the disorder parameter $W$
for $\theta = \pi/6$.
It is clear that the Anderson transition takes place at
$W_{\rm c}/V=19.0 \pm 0.2$ for this $\theta$.
The standard analysis for
the critical exponent $\nu$ of the localization length
defined as $ \xi(W) \sim |W-W_{\rm c}|^{-\nu}  ( W \rightarrow W_{\rm c}) $
yields $\nu = 1.3 \pm 0.2$. This value of $\nu$ is
close to those obtained for the orthogonal and the unitary symmetry
classes in 3D
\cite{KM,OKO,HKO,CD}.
The value of the renormalized localization length
at
the critical point $\Lambda_{\rm c}$ is
$\Lambda_{\rm c} = 0.56 \pm 0.02$ (Fig. 1), slightly smaller than
those obtained for the orthogonal and the unitary classes \cite{KM,HKO}.
A similar calculation for $\theta = \pi /4$ leads to $W_{\rm c}/V
=19.9 \pm 0.2$ with $\nu = 1.3 \pm 0.2$ and $\Lambda_{\rm c} =
0.55 \pm 0.02$. The fact that both $\nu$ and $\Lambda_{\rm c}$ are
insensitive to the strength of the spin-orbit coupling $\theta$
supports the single-parameter scaling behavior of $\xi_M/M$.
By introducing the scaling parameter $\xi$, we have
demonstrated that $\xi_M/M$ is indeed
a single parameter function of $M/\xi$
(Fig. 1, inset).

Now we focus on the level statistics at the
critical point.
Clarifying the relationship between the statistics at the
critical point and the critical behavior of the localization
length and of the conductivity is one of the
important open problems in the theory of the Anderson transition.
It has been proposed in refs.\cite{KLAA} and \cite{AKL} that the
asymptotic form of the critical spacing distribution $P_{\rm c}(s)$ of
neighboring energy levels is related to the
critical exponent $\nu$.
The proposed form for $P_{\rm c}(s)$ is
\begin{equation}
 P_{\rm c}(s) \sim \exp( -C s^{2-\gamma}), \quad s \gg 1,
\label{sbig}
\end{equation}
where $\gamma$ is related to the critical exponent $\nu$ of
the localization length
\begin{equation}
\gamma = 1- \frac{1}{d \nu}.
\label{gammatonu}
\end{equation}
Here $C$ is a positive constant, which may depend on the
symmetry of the system, and $d$ denotes the dimensionality.
For $s \ll 1$, the form of $P_{\rm c}(s)$ is expected to be
$P_{\rm c}(s) \sim s^{\beta}$, where $\beta$ is equal to $1,2$ and $4$
for the orthogonal,
the unitary and the symplectic universality classes, respectively.
It has also been suggested that the
overall shape of $P_{\rm c}(s)$ can be described by
the form \cite{AKL}:
\begin{equation}
 P_{\rm c}(s) = A s^{\beta} \exp(-B s^{2-\gamma}).
 \label{K-formula}
\end{equation}
The constants $A$ and $B$ are determined by the two constraints
$\int P(s) ds =1$ and $\int s P(s) ds =1$.
For a 3D system with  orthogonal symmetry
the above formula (\ref{K-formula}), with $\nu \approx 1.3$,
describes reasonably well
the numerical data for $P_{\rm c}(s)$
\cite{HS_3DO,Evangelou_3DO,VHSP}.

We have investigated the level spacing distribution $P_{\rm c}(s)$
at the critical point for our 3D symplectic system,
by numerically diagonalizing the Hamiltonians of
the $M \times M \times M$ systems with $M=6,8,10$ and $12$.
The numbers of realizations of random potentials are 2000 for $M=6$ and
1000 for $M=8,10$ and $12$.
One-tenth of the unfolded spectra around $E=0$ $(|E|
\buildrel{{\textstyle <}}\over{_{\textstyle \sim}}  V)$
is used for
calculating the spacing distribution $P_{\rm c}(s)$.
This is justified if $W_{\rm c}$ does not vary considerably
for energies $E$ close to the band center.
In fact, the critical disorder $W_{\rm c}$ for $E=2V$ and $\theta = \pi/6$
is estimated to be $W_{\rm c}/V= 18.9 \pm 0.5$,
which is almost the same as that for $E=0$.

The spacing distributions $P_{\rm c}(s)$
at the critical point $W_{\rm c}=19V(\theta = \pi /6)$
for $M=6$, 8, 10 and 12 are shown in Figs.
2 and 3. It is seen
that
$P_{\rm c}(s)$ is independent of
the size of the system $M$ confirming the existence of
a critical spacing distribution.

First we consider the limiting behavior of $P_{\rm c}(s)$
as $s\rightarrow 0$.
Here we find $P_{\rm c}(s) \sim s^{\beta}$ with
$\beta \simeq 4$ (see inset Fig. 2).
This quartic limiting behavior is
characteristic of systems with symplectic symmetry and
is a clear indication that the breaking of spin rotation
symmetry is relevant at the critical point.

Next we consider the behavior of $P_{\rm c}(s)$ in
the limit $s \gg 1$.
A fit to (\ref{sbig}) for $s>2$ yields $\gamma =1.0 \pm 0.15$.
Making use of (\ref{gammatonu}) this yields $\nu > 2.2$
which is inconsistent with the value of $\nu = 1.3 \pm 0.2$
we obtained earlier.
In fact the data are well fitted by a simple exponential
law  $P_{\rm c}(s) \sim \exp(- \alpha s)$ with
$ \alpha \approx 1.7$ (Fig. 2, solid line).
We note that a good fit to
data for 3D orthogonal \cite{ZK} and
2D symplectic systems \cite{SZ,Evangelou_2DS}
has also been obtained with a simple exponential.

Finally we consider a fit to the entire
spacing distribution.
If we take equation (\ref{K-formula}) with $\beta=4$
and $\gamma=0.75$, a value obtained by substituting
$\nu \simeq 1.3$ into (\ref{gammatonu}), we find a very poor
fit to $P_{\rm c}(s)$ (dashed line Fig 2).
On the other hand if we vary $\gamma$ so as to obtain
the best fit (see Fig 3) we find $\gamma=1.43 \pm 0.01$ which
corresponds
to $\nu < 0$ which is physically unacceptable.
We conclude that the proposed form, corresponding
to equations (\ref{gammatonu}) and (\ref{K-formula}),
is not consistent with our numerical data.
For comparison, we also plot in Fig. 3
$P_{\rm c}(s)$ for
a 2D symplectic system \cite{SZ,OO,Evangelou_2DS} where the
exponent $\nu$ is estimated to be $\nu \approx 2.7$ \cite{FABHRSWW}.
It is worth noting that although
the estimated critical exponent $\nu$ in 2D is roughly twice our
estimate for $\nu$ in 3D
the corresponding values of $\gamma$ obtained from (\ref{gammatonu})
are close to each other.
If the formulae (\ref{gammatonu}) and (\ref{K-formula}) were valid,
$P_{\rm c}(s)$ in 2D would have to look similar to that in the
3D system.
We see in Fig. 3 that this is clearly not so.

In summary, we have analyzed the Anderson transition in a
3D system with symplectic symmetry. The
critical exponent $\nu$ of the localization length is estimated
to be $\nu = 1.3 \pm 0.2$,
which is rather close to
the values found in 3D unitary and orthogonal systems.
On the other hand, we have demonstrated that the
energy level spacing distribution $P_{\rm c}(s)$
at the critical point is sensitive to
the breaking of spin rotation symmetry.
In particular for $s\ll 1$ we find $P_{\rm c}(s) \sim s^{\beta}$
with $\beta \simeq 4$ which is characteristic of the
symplectic symmetry class.
The sensitivity of $P_{\rm c}(s)$, at small $s$, to the
breaking of  spin rotation symmetry is consistent with the
conventional classification of the critical behavior
according to symmetry.

One of the authors (T.K.) thanks M. Takahashi for discussions.
The numerical calculations were done on a FACOM VPP500 of
Institute for Solid State Physics, University of Tokyo.
This work was partly supported by the Grants-in-aid No. 07740321,
No. 07640520 and No. 07740334
for Scientific Research from the Ministry of Education, Science and
Culture, Japan.


%
%
\begin{figure}
\caption{The renormalized localization length $\Lambda_M$ as a function
of the disorder $W$ for $\theta = \pi /6$. Open triangles, open diamonds,
open squares and open circles correspond to
$M=6$, 8, 10 and 12, respectively.
Inset: One-parameter scaling behavior of $\Lambda_M$ for $\theta = \pi /6$
and $\pi /4$. Filled triangles, filled diamonds, filled squares and
filled circles represent the data in the case of $\theta  = \pi /4$
for $M=6$, 8, 10 and 12, respectively, while open marks are for
$\theta = \pi /6$.}
\end{figure}
\begin{figure}
\caption{The spacing distribution
$P_{\rm c}(s)$ at the critical point $W=W_{\rm c}
(=19V)$ for $\theta = \pi /6$. Open triangles, open diamonds, open squares
and open circles stand for the data for $M=6$, 8, 10 and 12, respectively.
Dashed curves represent the result by the
formula (\protect\ref{K-formula})
with $\gamma = 0.75$ and $\beta =4$ . Solid line is
a function $ \propto \exp(-1.68 s)$.
Inset: Double logarithmic plot of
$P_{\rm c}(s)$ for $s \ll 1$. The solid line expresses
the behavior $\propto s^4$ whereas the dashed lines express
$\propto s$.}
\end{figure}
\begin{figure}
\caption{The spacing distribution $P_{\rm c}(s)$
at the critical point $W=W_{\rm c}
(=19V)$ for $\theta = \pi /6$ is shown in the linear-scale.
For comparison,
the data of the critical spacing distribution for the 2D symplectic system,
which is taken from Ref.\protect\cite{OO},
are also plotted (crosses).
The dashed curve represents the Wigner surmise for the Gaussian
symplectic ensemble.
The solid one represents the best fit based on
the formula (\protect\ref{K-formula})
with $\gamma=1.43$ and $\beta=4$.}
\end{figure}

\end{document}